\documentclass[onecolumn,showpacs,preprint,preprintnumbers,amsmath,amssymb]{revtex4}
\usepackage{amssymb}
\usepackage[dvips]{graphicx}
\begin{document}

\title{ Phase coexistence and  resistivity  near the ferromagnetic transition of manganites}
\author{A. S. Alexandrov$^{1}$, A.M. Bratkovsky $^{2}$ and V.V. Kabanov $^{3}$}

\affiliation{$^{1}$ Department of Physics, Loughborough University, Loughborough LE11 3TU, United Kingdom\\
$^2$Hewlett-Packard Laboratories, 1501 Page Mill Road, 1L, Palo
Alto, California 94304\\
$^{3}$ Josef Stefan Institute 1001, Ljubljana, Slovenia}

\begin{abstract} Pairing of oxygen holes into heavy bipolarons in the paramagnetic phase and
their magnetic pair-breaking in the ferromagnetic phase [the
so-called current-carrier density collapse (CCDC)] has accounted for
the first-order ferromagnetic phase transition, colossal
magnetoresistance (CMR), isotope effect, and pseudogap in doped
manganites. Here we propose an explanation of the phase coexistence
and describe the magnetization and resistivity of manganites near
the ferromagnetic transition in the framework of CCDC. The present
quantitative description of resistivity is obtained without any
fitting  parameters by using the experimental resistivities  far
away from the transition and the experimental magnetization,  and
essentially model independent.
\end{abstract}


\maketitle

  Ferromagnetic oxides, in particular manganese perovskites,
show very large magnetoresistance near the ferromagnetic transition.
The effect observed in these materials was termed ``colossal''
magnetoresistance (CMR) to distinguish it from the giant
magnetoresistance in metallic magnetic multilayers. The discovery
raised expectations of a new generation of magnetic devices, and is
a focus of extensive research aimed at describing the effect.
Significant progress has been made in understanding the properties
of CMR\ manganites, but many questions remain. The ferromagnetic
metal-insulator transition in manganites has long been thought to be
a consequence of the so-called double exchange mechanism (DEX),
which results in a varying bandwidth of electrons in the Mn$^{3+}$
d-shell as a function of temperature \cite{dex}. More recently, it
has been noticed \cite{mil} that the effective spin interaction
cannot alone account for CMR within the double-exchange model. In
fact, there is strong experimental evidence for exceptionally strong
electron-phonon interactions in doped manganites from the giant
isotope effect \cite{zhao1}, the Arrhenius behaviour of the drift
and Hall mobilities \cite{emi0} in the paramagnetic phase above the
Curie temperature, $T_{C}$, and other experiments. In view of this,
Ref. \cite{mil} and some subsequent theoretical studies have
combined DEX with the Jahn-Teller e-ph interaction with d-states
arriving at the conclusion that the low-temperature ferromagnetic
phase is a spin-polarised metal, while the paramagnetic
high-temperature phase is a polaronic insulator.

   However, some low-temperature optical \cite{opt},
electron-energy-loss (EELS) \cite{eels}, photoemission \cite{arpes}
and thermoelectric \cite{thermo} measurements showed that the
ferromagnetic phase of manganites is not a conventional metal. In
particular, broad incoherent spectral features and a pseudogap in
the excitation spectrum were observed. EELS confirmed that
manganites were in fact charge-transfer doped insulators having
p-holes as current carriers, rather than d(Mn$^{3+})$
electrons. Photoemission and x-ray absorption spectroscopies of La$_{1-x}$Sr$%
_{x}$MnO$_{3}$ also showed that the itinerant holes doped into
LaMnO$_{3}$ have oxygen p-character. Further, CMR has been observed
in the ferromagnetic pyrochlore manganite Tl$_{2}$Mn$_{2}$O$_{7}$
\cite{ram}, which has neither the mixed valence for DEX magnetic
interaction nor the Jahn-Teller cations such as Mn$^{3+}$.

   \begin{figure}
\begin{center}
\includegraphics[angle=-0,width=0.60\textwidth]{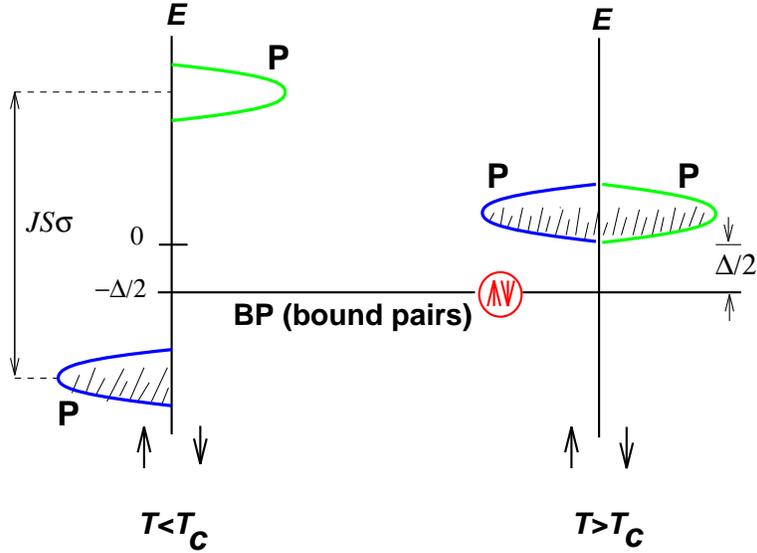}
\vskip -0.5mm \caption{Bipolaron model of CMR:  pairs (BP) are
localised on  impurity levels in the paramagnetic phase, where the
only current carriers are single thermally excited polarons. If the
exchange interaction $JS\sigma$ between p-hole polarons and ordered
manganese spins exceeds the pair binding energy $\Delta$ , the pairs
break at $T<T_C$ because  the spin-up polaron sub-band sinks
abruptly below the bipolaron level. The ferromagnetic state is a
\emph{polaronic} conductor.}
\end{center}
\end{figure}

These and other observations \cite{zhao2}, in particular the fact
that some samples of ferromagnetic manganites manifest an
insulating-like optical conductivity at all temperatures \cite{emi},
clearly rule out DEX as the mechanism of CMR. The earlier of the
above observations \cite{zhao1,emi0,opt,eels,ram}  led to a novel
theory of ferromagnetic/paramagnetic phase transition and CMR, based
on the so-called current-carrier density collapse (CCDC)
\cite{alebra2}, confirmed by later observations.  In CCDC model the
p-holes are bound into heavy bipolarons above the Curie temperature
$T_{C}$ due to the Fr\"{o}hlich electron-phonon interaction, which
is written in the real-space representation as
\begin{equation}
H_{e-ph}=\sum_{\mathbf{n}\mathbf{n'}\sigma=\uparrow, \downarrow}
f_{\mathbf{n'}}(\mathbf{n}) c^{\dagger}_{\mathbf{n}\sigma}
c_{\mathbf{n}\sigma} \xi_{\mathbf{n'}},
\end{equation}
where $\xi_{\bf n'}$ is the ion displacement  operator, and the form
of electron-phonon interaction is specified via the {\em force}
function \cite{alebook} $f_{\bf n'}({\bf n})$.  The latter is
defined as the force with which an electron in state $\vert {\bf n}
\rangle$ interacts with the ion degree of freedom $\xi_{\bf n'}$.

 The resistivity peak and
CMR are the result of the magnetic pair-breaking below $T_{C}$,
Fig.~1, caused by the $p-d$ spin-exchange interaction, $J_{pd}$,
described as
\begin{equation}
H_{pd}= -(2N)^{-1} \sum_{\bf n,m} J_{pd}\hat{S}^z_{\bf
m}(c^{\dagger}_{\bf n \uparrow}c_{\bf n \uparrow}-c^{\dagger}_{\bf n
\downarrow}c_{\bf n \downarrow}).
\end{equation}
Here $\hat{S}^z_{\bf m}$ is the z-component of $Mn^{3+}$ spin on
site ${\bf m}$, $c_{\bf n \uparrow}$ and $c_{\bf n \downarrow}$
annihilate a p($\uparrow, \downarrow$)-hole on the oxygen site, $\bf
n$, with spin up and down, respectively, and $N$ is the total number
of unite cells.

Different from cuprates,  hole  on-site  or more extended intersite
oxygen bipolarons are much heavier in manganites because the e-ph
Fr\"{o}hlich interaction, Eq.(1), is stronger \cite {alebra} and the
band structure is less anisotropic. They are readily localised by
disorder, so it is mainly thermally-excited single  polarons that
conduct in the paramagnetic phase. Upon temperature lowering single
polarons polarize manganese spins at $T_{C}$ via the exchange
interaction $J_{pd}$, and the spin polarization
of manganese ions breaks the bipolaronic singlets creating a spin-polarized $%
polaronic$ conductor.

The CCDC model has explained CMR in the experimental range of
external magnetic fields \cite{alebra2,tai}. More recently, the
theory has been further confirmed experimentally. In particular, the
oxygen isotope effect has been observed in the low-temperature
resistivity of La$_{0.75}$Ca$_{0.25}$MnO$_3$ and
Nd$_{0.7}$Sr$_{0.3}$MnO$_3$ and explained by CCDC with polaronic
carriers in the ferromagnetic phase \cite{alezhao}. The
current-carrier density collapse has been directly observed using
the Hall data in La$_{0.67}$Ca$_{0.33}$MnO$_3$ and
La$_{0.67}$Sr$_{0.33}$MnO$_3$ \cite{hall}, and the first order phase
transition at $T_C$, predicted by  the theory \cite{alebra2}, has
been firmly established in the specific heat measurements
\cite{phil}. Importantly the character of the magnetic phase
transition in Tl$_2$Mn$_2$O$_7$ pyrochlores has also been determined
to be the first order \cite{gui1} and attributed to the tendency of
small polarons to phase separation at finite carrier density. Indeed
recent Monte Carlo simulations \cite{kab0} of lattice polarons with
anisotropic e-ph interactions and the realistic long-range Coulomb
repulsion show diverse mesoscopic textures in the adiabatic limit,
where spatially disordered \emph{pairs} (i.e. bipolarons) dominate
at finite doping.

On the other hand,  resistivity and the magnetization of some
La$_{0.7}$Ca$_{0.3}$Mn$_{1-\delta}$Ti$_{\delta}$O$_{3}$ samples
showed a more gradual (second-order like) transition \cite{china}.
Also the coexistence of ferromagnetic and paramagnetic phases near
the Curie temperature observed in tunnelling \cite{tun} and  other
experiments has not yet been  addressed in the framework of CCDC.
Here we show that the diagonal disorder, which is inevitable with
doping in those solid-state solutions, explains both the phase
coexistence and the resistivity/magnetization behaviour near the
transition.

The mean-field equations \cite{alebra2} describing  p-hole polaron
atomic density, $n$,  polaron, $m$, and manganese, $\sigma$,
 reduced magnetizations, and the chemical potential $\mu=k_BT \ln y$ are readily
generalized taking into account a random distribution of the
bipolaron binding energy $\delta= \Delta/(2J_{pd})$ across the
sample,
\begin{eqnarray}
&&n_i= 6y \cosh (\sigma_i/t), \cr
&& m_i= n_i \tanh (\sigma_i/t), \cr
&& \sigma_i= B_2 (m_i/2t), \cr
&& y^2= {{x-n_i}\over{18}}\exp(-2\delta_i/t),
\end{eqnarray}
where $t=k_B T/J_{pd}$ is the reduced temperature, $B_S$ is the
Brillouin function, $x$ is the number of  holes at zero temperature
in p-orbital states, which are   $3$-fold degenerate. The subscript
$i$ means different parts of the sample with different $\delta_i$
and hence \cite{alebra2}  with different Curie temperatures,
$T_{Ci}$, owing to disorder.

\begin{figure}
\begin{center}
\includegraphics[angle=-90,width=0.60\textwidth]{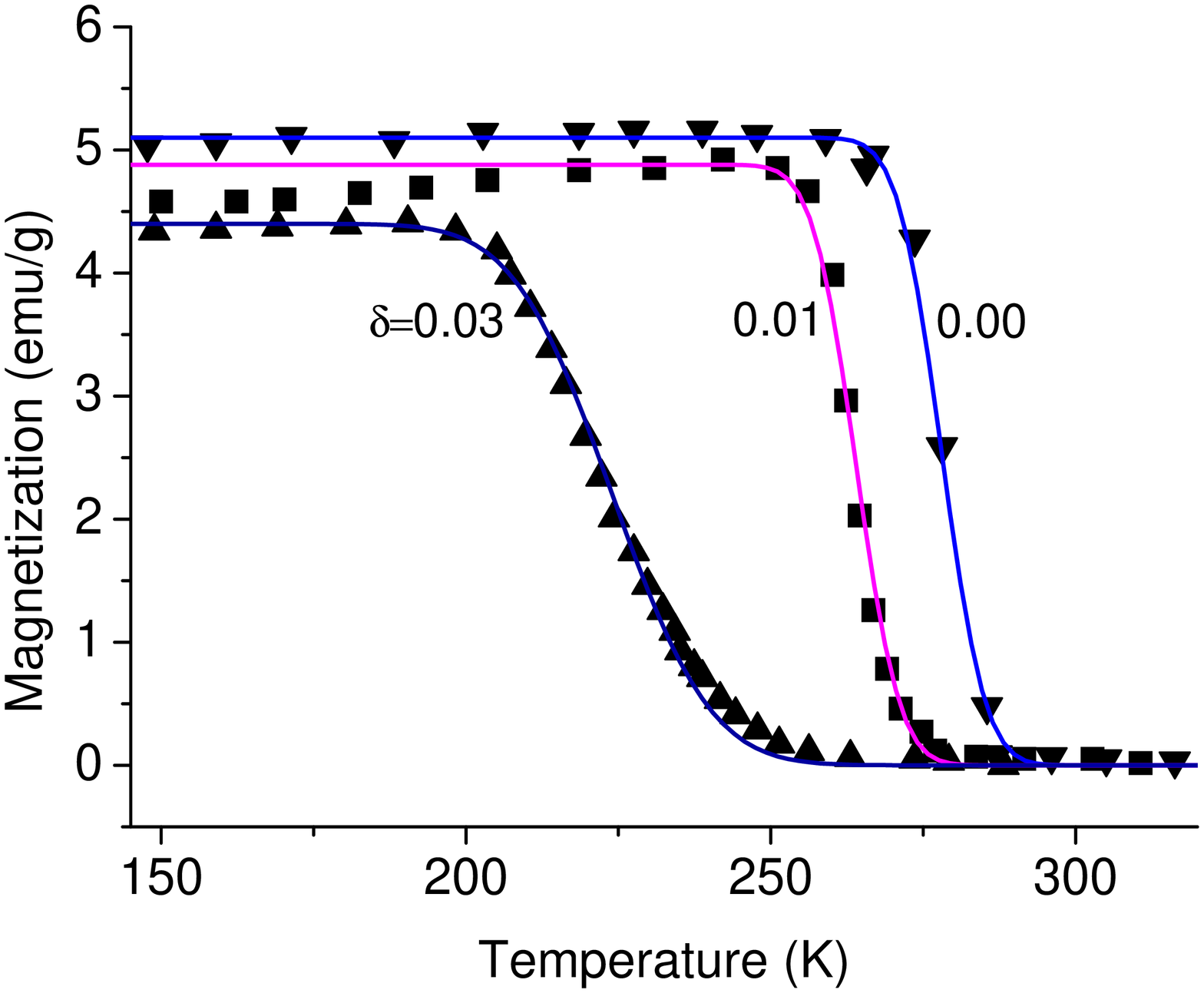}
\vskip -0.5mm \caption{ Experimental magnetizations  in
La$_{0.7}$Ca$_{0.3}$Mn$_{2-\delta}$Ti$_{\delta}$O$_{3}$ (symbols
\cite{china}) compared with Eq.(4)(lines) with $\Gamma=8$K,
$T_C=278$K for $\delta=0.00$; $\Gamma=8$K, $T_C=264$K for
$\delta=0.01$; and $\Gamma=18$K, $T_C=224$K for $\delta=0.03.$}
\end{center}
\end{figure}

While averaging these simple equations over a random distribution of
$\delta_i$ is rather cumbersome, one can apply a simplified approach
using the fact that the  phase transition in a homogeneous system is
of the first order in a wide range of $\delta$ \cite{alebra2}.
Taking $\sigma_i\approx \Theta(T_{Ci}-T)$ and $n_i\approx x
\Theta(T_{Ci}-T)+ \sqrt{2x} \exp(-\Delta/(2k_BT))\Theta(T-T_{Ci})$,
and averaging both quantities with the Gaussian distribution of
random $T_{Ci}$s around the experimental $T_C$ we obtain an averaged
manganese magnetization
\begin{equation}
\sigma(T)={1\over{2}} \mathrm{erfc}
\left({{T-T_C}\over{\Gamma}}\right),
\end{equation}
where $\Delta$ is the average bipolaron binding energy,
$\Theta(y)=1$ for $y>0$ and zero for $y<0$ , and $\mathrm{erfc}(z)=
(2/\pi^{1/2}) \int _z^{\infty} dy \exp(-y^2)$.
 The CCDC with disorder, Eq.~(4)
fits nicely the experimental magnetizations \cite{china} near the
transition with physically reasonable $\Gamma $ of the order of 10K,
depending on doping, Fig.~2. Hence, we believe that the random
distribution of transition temperatures with the width $\Gamma $
across the sample caused by the randomness of the bipolaron binding
energy is responsible for the phase coexistence near the transition
as seen in the tunnelling experiments \cite {tun}. We note that some
drop of magnetization at low temperatures as seen in Fig.~2 might be
caused by domain walls \cite{sench}.

Resistivity of inhomogeneous two-phase systems has to be calculated
numerically. Nevertheless, the comprehensive numerical simulations
are consistent with a simple analytical expression for the
resistivity of the binary mixture,
\begin{equation}
\rho= \rho_1^{1-\nu}\rho_2^{\nu},
\end{equation}
which is valid in a wide range of the ratios $\rho _{1}/\rho _{2}$
\cite {kab}. Here
 $\rho_{1,2}$ is the resistivity of each phase, respectively,
and $\nu$ is the volume fraction of the second phase. The expression
Eq.(5)
 is a homogeneous
function of $\rho_{1,2}$   satisfying  the duality relation.
 Scaling arguments \cite{eff} proved that the expression
is exact near a percolation threshold  in two dimensions. Numerical
analysis \cite{kab}  has shown that Eq.(5) describes the effective
resistivity of 2D random systems even far  away from the percolation
threshold, if the resistivity ratio $\rho_1/\rho_2$ is not extremely
large ($\lesssim 20$). The same expression is also asymptotically
correct for any 3D system, if $|1-\rho_1/\rho_2|\ll 1$, and  for
specific 3D lattice structures \cite{kel},  if $\rho_1/\rho_2\gg 1$.
Of course, there is no universal formula for any material with
randomly distributed phases. Generally, one could write
$f(\rho)=(1-\nu) f(\rho_1)+\nu f(\rho_2)$, where $f(x)$ is a model
function (for a comprehensive list of mixture formulae see
Ref.\cite{milt,shi}). Equation (5) corresponds to the average  of
$\ln{\rho}$ in isotropic mixtures providing a qualitatively
reasonable and numerically accurate description of the effective
resistivity in many physically important cases (see below).

\begin{figure}
\begin{center}
\includegraphics[angle=-0,width=0.80\textwidth]{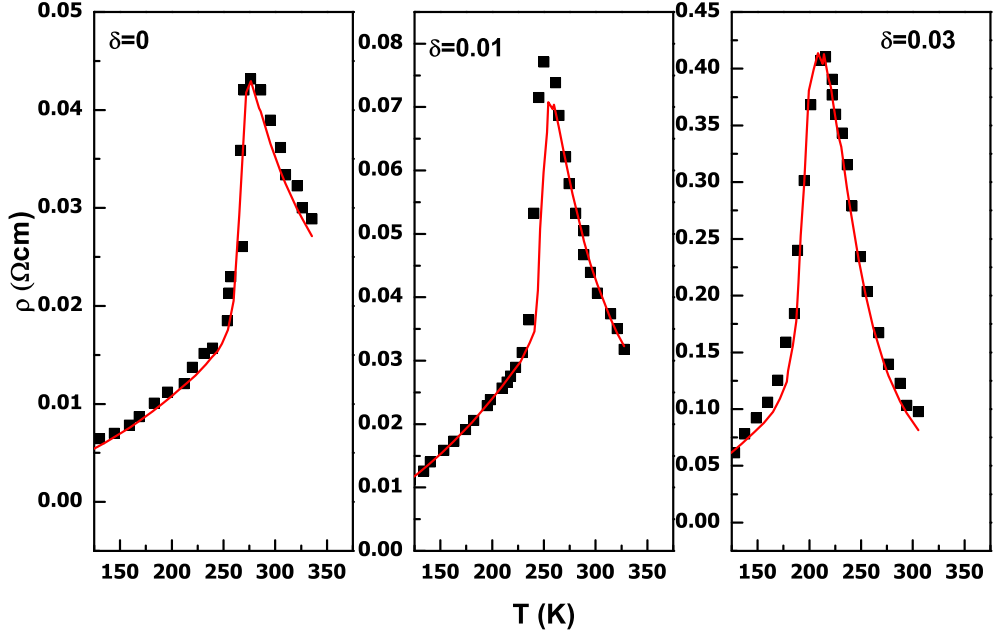}
\vskip -0.5mm \caption{CCDC model (Eq.(5), lines) describes the
experimental
 resistivity near the ferromagnetic transition
in La$_{0.7}$Ca$_{0.3}$Mn$_{2-\delta}$Ti$_{\delta}$O$_{3}$ (squares
\cite{china}), if the phase coexistence caused by disorder is taken
into account. No fitting parameters are used in Eq.(5) but the
experimental resistivity well below and well above the transition
and the experimental magnetization. }
\end{center}
\end{figure}

In the framework of CCDC, the resistivity of the paramagnetic phase
is $\rho _{1}(T)=f(T)\exp (\Delta /2k_BT)$ and the resistivity of
the ferromagnetic phase is $\rho _{2}(T)=\phi (T)$, where $f(T)$ and
$\phi (T)$ are polynomial functions of temperature depending on the
scattering mechanisms. Well below the transition $\phi (T)$ can be
parameterized  as $\phi (T)=\rho _{0}+aT^{2}$, and $f(T)=bT$ well
above the transition, where the temperature independent parameters
$\rho _{0}$, $a$, $\Delta/2$ and $b$ are taken directly from the
experiment \cite{china}. The microscopic origin of $\rho_0$, $a$,
$b$ and alternative parametrization formulae have been discussed
e.g. in Ref.\cite{zhao3,china}), and are not an issue here.
The volume fraction $%
\nu $ of the ferromagnetic phase is simply the relative
magnetization in our model, $\nu =\sigma (T)$, also available from
the experiment \cite{china}. As a result, Eq.~(5) provides the
quantitative description of $\rho (T)$ in the  transition region
without any fitting  parameters by using the experimental
resistivities  far away from the transition and the experimental
magnetization \cite{china},  as shown in Fig.3. On the other hand,
fitting the ferromagnetic-phase resistivity with a magnetic
scattering ($\rho
_{4.5}^{m}T^{4.5}$) leads to an unrealistic doping-dependent coefficient $%
\rho _{4.5}^{m}$ that is changing with doping by more than five
orders of magnitude (see the Table in Ref. \cite{china}). Note, that
if one were using an estimate, $\rho \propto 1/n$, where $n$ is the
average single-polaron density, one would obtain
\begin{equation}
\rho^{-1}(T)\propto \mathrm{erfc}\left( {\frac{{T-T_{C}}}{{\Gamma
}}}\right) +(2/x)^{1/2}e^{-\Delta /2k_BT}\mathrm{erfc}\left(
{\frac{{T_{C}-T}}{{\Gamma }}}\right) .
\end{equation}
This expression can also fit the experimental curves, but with a value of $%
T_{C},$ which turns out to be smaller than that in the
magnetization, Eq.~(4), by several tens of degrees Kelvin
\cite{ale}. The latter expression corresponds to a linear expansion
of Eq.~(5) in powers of $1-\rho _{1}/\rho _{2}.$ It is easy to see
why Eq.~(5), when compared with Eq.~(6), resolves the problem of
different $T_{C}$s in the magnetization and resistivity, thus
providing a parameter-free description of experimental $\rho (T)$.
If we take $\rho _{1}\gg \rho _{2}$, the resistivity at the magnetic
transition (i.e. for $\nu =\sigma =1/2$), $\rho =\rho _{2}\sqrt{\rho
_{1}/\rho _{2}}$, calculated with Eq.~(5) turns out larger than the
resistivity $\rho \approx 2\rho _{2}$ calculated with the
perturbation expression, Eq.~(6). It is important that the present
description of resistivity does not depend on a particular model,
but on the assumption that the ferromagnetic transition is the first
order. The CCDC \cite{alebra2} has provided a basis for this
assumption in terms of the microscopic model.

 In summary, we have shown that the conventional DEX model,
proposed half a century ago and generalized more recently to include
the electron-phonon interaction, is in conflict with a number of
recent experiments. Among these experiments are the site-selective
spectroscopies, which have shown unambiguously that oxygen p-holes
are the current carriers rather than d-electrons in ferromagnetic
manganites. Also, some samples of ferromagnetic manganites manifest
an insulating-like optical conductivity at all temperatures,
contradicting the DEX notion that their ferromagnetic phase is
metallic. On the other hand, the pairing of oxygen holes into heavy
bipolarons in the paramagnetic phase and their magnetic break-up in
the ferromagnetic phase has explained the colossal
magnetoresistance, the isotope effects, and the pseudogaps observed
in doped manganites. It also explains the CMR \ in systems where
DEX\ simply cannot exist, like manganese pyrochlores  \cite{ram}.
The CCDC theory of CMR predicts the first-order phase transition and
allows the present simple explanation of the coexistence of high and
low-resistive phases. It explains the temperature dependencies of
the magnetization and the resistivity near the transition as the
result of the unavoidable disorder and transport through the
two-phase mixture in doped manganites.

  This work was supported by  EPSRC  (UK) (grants
EP/D035589/1 and EP/C518365/1).

\end{document}